\documentclass{pasa}%

\usepackage{graphicx,xcolor}
\usepackage{aas_macros}
\usepackage{hyperref} 
\hypersetup{colorlinks,citecolor=blue,linkcolor=blue,urlcolor=blue}

\title[Exomoons in HD~23079]{Updated Studies on Exomoons in the HD~23079 System}

\author[Jagtap et al.]{O. Jagtap$^{1}$,
B. Quarles$^{2,3}$\thanks{Corresponding Author}, and M. Cuntz$^{1}$
\affil{$^1$Department of Physics, University of Texas at Arlington, Arlington, TX 76019, USA}%
\affil{$^2$Center for Relativistic Astrophysics, School of Physics, Georgia Institute of Technology, Atlanta, GA 30332, USA}
\affil{$^3$Department of Physics, Astronomy, Geosciences and Engineering Technology, Valdosta State University, Valdosta, GA 31698, USA}
}%

\jid{PASA}
\jyear{\the\year}

\hypersetup{draft}

\begin{document}

\begin{frontmatter}
\maketitle

\begin{abstract}
We re-evaluate the outer edge of orbital stability for possible exomoons orbiting the radial velocity planet discovered in the HD~23079 system.  In this system, a solar-type star hosts a Jupiter-mass planet in a nearly circular orbit in the outer stellar habitable zone.  The outer stability limit of exomoons is deduced using $N$-body and tidal migration simulations considering a large range of initial conditions, encompassing both prograde and retrograde orbits.  In particular, we extend previous works by evaluating many values in the satellite mean anomaly to identify and exclude regions of quasi-stability. Future observations of this system can make use of our results through a scale factor relative to the currently measured minimum mass. Using a constant time lag tidal model \citep{Hut1981}, we find that plausible tidal interactions within the system are insufficient to induce significant outward migration toward the theoretical stability limit.  While current technologies are incapable of detecting exomoons in this system, we comment on the detectability of putative moons through Doppler monitoring within direct imaging observations {in view of future research capacities}.

\end{abstract}

\end{frontmatter}

\section{INTRODUCTION }
\label{sec:intro}
The detection of HD~23079b, reported by \cite{Tinney2002}, was a successful outcome of the {\it Anglo-Australian Planet Search}.  HD~23079b, a Jupiter-type planet, is hosted by a solar-type star with a temperature of about 6000~K \citep{Bonfanti2015}; see Table~1 for details.  This system is located in the Southern sky in the constellation Reticulum.  HD~23079b is in a nearly circular orbit situated at the outskirts of the stellar habitable zone (HZ) that extends between 0.87 and 2.03~au {\citep[optimistic limits; see][]{Kopparapu2013,Kopparapu2014}}.

The relatively low level of stellar activity of HD~23079, consistent with its age of $\sim$5~Gyr (see Sect. 2.1), tends to favor the existence of a habitable circumstellar environment; see, e.g.,
\cite{Ribas2005,Lammer2009,Ramirez2018} for more general discussions.
However, massive planets such as HD~23079b with orbits located within stellar HZ tend to thwart the existence of habitable terrestrial planets owing to the onset of orbital instabilities {\citep[e.g.,][]{Jones2001,Noble2002,Agnew2017,Agnew2018}}.
Nevertheless, there is a significant possibility for the existence of habitable Trojan planets and/or habitable exomoons (in orbit about HD~23079b), as demonstrated via detailed simulations; see \cite{Eberle2011} and \cite{Cuntz2013}, respectively.

The search for exomoons has been an active endeavor after the launch of the {\it Kepler Space Telescope}, while many works \citep{Sartoretti1999,Cabrera2007,Kipping2009a,Kipping2009b} preceding the {\it Kepler} era laid the theoretical groundwork for their detection through transit timing and duration variations.  However, such methods have limitations \citep{Kipping2020,Kipping2021} and photometric observations can still lead to false positives, including one candidate for Kepler-90g \citep{Kipping2015a}.

Fortunately, there are other methods proposed for exomoon detection including using {a planet profile determined by the average light curve} \citep{Simon2012}, optimizing with respect to the orbital sampling effect \citep{Heller2014,Heller2016,Hippke2015}, Doppler monitoring of directly imaged exoplanets \citep{Agol2015,Vanderburg2018}, or examining the radio emissions from giant exoplanets \citep{Noyola2014,Noyola2016}.  Another motivation for our study stems from the recent discovery of a circumplanetary disk (system PDS 70), indicating the ongoing formation of one or more exomoons in alignment with the Hill radius criterion \citep{Benisty2021}.

Theoretical constraints aid in the interpretation of observations and can be useful to quickly validate whether a exomoon candidate is plausible or not \citep{Quarles2020b}.  One of these constraints is the combined tidal interaction between the host star, planet, and moon \citep{Barnes2002,Sasaki2012,Sasaki2014,lainey2020resonance} that generally depends on a wide range of parameters (e.g., tidal Love number and tidal quality factor).  \cite{Spalding2016} explored how the so-called {`evection resonance' can cause} significant growth in a moon's eccentricity, which can lead to the moon's tidal breakup or escape from the planet's gravitational influence.

Nearby \citep{Payne2013} and distant planetary companions \citep{Grishin2017} can also {drive an exomoon along a similar path to destruction.}  Even without these confounding interactions, \cite{Domingos2006} produced estimates for exomoon stability using three-body interactions, but these results represent the upper boundary of a transition region for stability \citep{Dvorak1986}.  Recently, \cite{Rosario-Franco2020} determined a revised fitting formula for the (more conservative) lower stability boundary for prograde satellites, whereas \cite{Quarles2021} derived a similar fitting formula for retrograde satellites.

In this study, we revisit the existence of possible exomoons in the HD~23079 system based on more generalized assumptions and an improved methodology.  Our paper is structured as follows.  In Section 2, we summarize our theoretical approach.  Our results and discussion are conveyed in Section 3 including comparisons to previous works.  Here we also comment on the observability of possible HD~23079 exomoons.  In Section 4, we report our summary and conclusions.

\section{THEORETICAL APPROACH}
\subsection{Stellar and planetary parameters}
HD~23079 is a solar-type star of spectral type F9.5V \citep{Gray2006} with an effective temperature of about 6003~K \citep{Bonfanti2015}; see Table~1.  Its mass and radius are given as 
$1.01 \pm 0.02$ M$_\odot$ and $1.08 \pm 0.02$ R$_\odot$, respectively.
HD~23079 has an age of approximately 5~Gyr \citep{Saffe2005,Bonfanti2015}, which implies a relatively low level of chromospheric activity --- a notable feature in support of circumstellar habitability \citep[e.g.,][]{Kasting2003,Lammer2009,Kaltenegger2017}.
The minimum mass $m_{\rm p}{\sin}i$ of the planet HD~23079b, discovered by \cite{Tinney2002}, has been identified as $2.41 \pm 0.6$ M$_{\rm J}$; however, the exact value of $m_{\rm p}$ is unknown owing to the inherent limitations of the Radial Velocity (RV) method.

The stellar luminosity is about 35\% larger than that of the Sun;
hence, the HZ of HD~23079 is notably wider and further extended than the {Solar} HZ.  In fact, the outer limits of the conservative and optimistic HZ are identified as 1.93 and 2.03~au, respectively.
The orbital parameters of the planet, i.e., the semimajor axis 
$a_{\rm p}$ and the eccentricity $e_{\rm p}$, are given as
1.586 $\pm$ 0.003 au and 0.087 $\pm$ 0.031, respectively, indicating that HD~23079b is situated in a nearly circular orbit within the stellar HZ at an orbital distance akin to that of Mars relative to the Sun.  The planetary Hill radius is given as:

\begin{equation} \label{eqn:R_H}
    R_{\rm H} = a_{\rm p}\left(\frac{m_{\rm p}+m_{\rm sat}}{3M_\star}\right)^{1/3},
\end{equation}
which includes the planet, satellite, and stellar mass ($m_{\rm p}$, $m_{\rm sat}$, and $M_\star$, respectively) in addition to the planetary semimajor axis $a_{\rm p}$.   {In physical units, the Hill radius is approximately 0.144 au using the appropriate values from Table \ref{tab:sys_param}.}  This formulation of the Hill radius is appropriate because the planetary eccentricity is low and no significant third body exists that can substantially force the planetary eccentricity \citep{Quarles2021}.

\subsection{$N$-body simulations} \label{sec:nbody}

To investigate the potential for exomoons in HD~23079, we perform a series of numerical simulations that identify the orbital stability of an Earth-mass satellite orbiting HD~23079b, a Jupiter-like planet. The numerical simulations are carried out using the general $N$-body software \texttt{REBOUND} \citep{Rein2012} with its \texttt{IAS15} adaptive step integration scheme \citep{Rein2015}.  The \texttt{IAS15} integrator is necessary because our study explores both prograde ($i_{\rm sat} = 0^\circ$) and retrograde ($i_{\rm sat} = 180^\circ$) satellite orbits, where the latter can be highly eccentric.  Adaptive step integrators, although more accurate, can also be more computationally expensive.

We set the initial timestep equal to 5\% of the shortest satellite orbital period ($\sim$0.007 yr for prograde or $\sim$0.017 yr for retrograde) and define 0.0001 yr as the minimum allowed timestep with the default accuracy parameter $10^{-9}$ used for the \texttt{IAS15} integrator.  \cite{Cuntz2013} showed that the outcomes of simulations are identical for timesteps smaller than the prescribed minimum using other adaptive timestep methods.  The simulation timescale of our $N$-body integrations is $10^5$ years, which is typical for determining the stability limits for hierarchical systems within large parameter spaces \citep{Rosario-Franco2020,Quarles2020a,Quarles2021}.

\begin{figure*} 
\centering
  \includegraphics[width=0.65\linewidth]{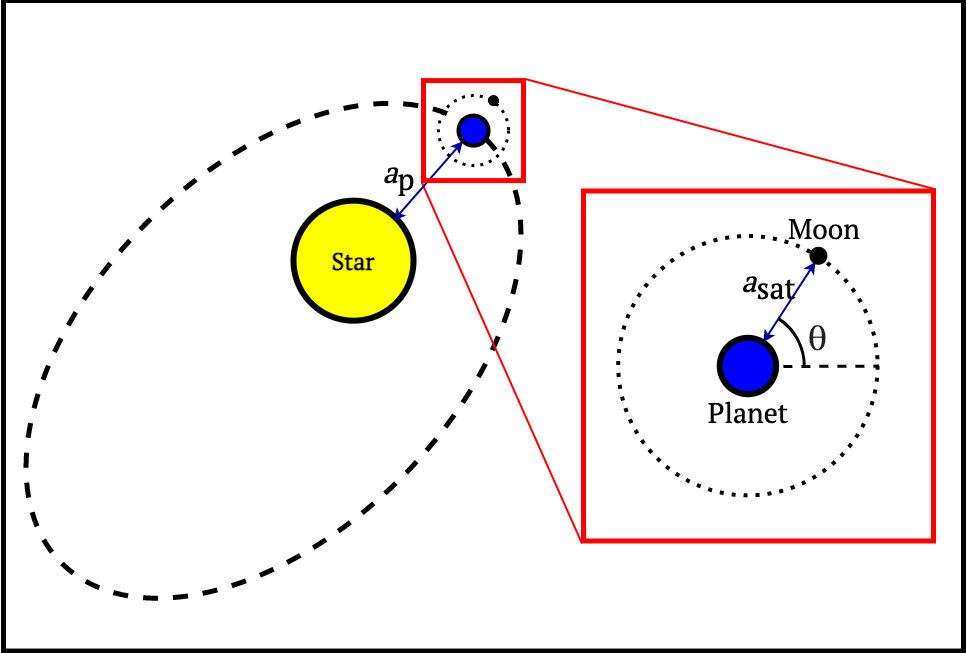}
  \caption{Representation of the initial conditions of the HD~23079 system in our calculations: (a) Planet HD~23079b (blue dot) orbits star HD~23079 (yellow dot). (b) An exomoon (black dot) orbits HD~23079b. Initially, the planet HD~23079b starts at perihelion and the exomoon starts at a random angle ($\theta$) with respect to the planet for each simulation. \label{fig:cartoon}}
  
\end{figure*}

Each simulation begins centered around the host star, HD~23079, with the host planet and satellite added hierarchically using a Jacobi coordinate system (see Figure \ref{fig:cartoon}).  The planet begins at its periastron position $\omega_{\rm p}$ and the line of apsides $\Omega_{\rm p}$ is used as the reference direction ($\omega_{\rm p} = \Omega_{\rm p} = 0^\circ$).  {An initial condition is classified as potentially stable if the satellite does not encounter either of our stopping criteria to detect instabilities. We stop our simulations and classify an initial condition as unstable} if the putative satellite: a) crosses the planet's Hill radius thereby {leaving the region over which the planet's gravitational influence dominates over that of the star} or b) collides with the host planet over a given timescale.  In addition, we require that a stable initial condition does not depend on the initial mean anomaly $\theta$ of the satellite (Fig. \ref{fig:cartoon}), which largely excludes islands of quasi-stability due to mean motion resonances \citep{Mudryk2006}.

\cite{Cuntz2013} explored a parameter space that varied the initial planetary semimajor axis, eccentricity, and the satellite's semimajor axis $a_{\rm sat}$.  Recent observations \citep{Wittenmyer2020} greatly narrowed the uncertainty of the planetary semimajor axis; therefore, we keep the planetary semimajor axis fixed ($a_{\rm p} = 1.586$ au) throughout this work.  However, we evaluate simulations varying the planetary eccentricity $e_{\rm sat}$ from 0.05 to 0.12 in 0.001 steps motivated by the observational uncertainties.  The prograde simulations are evaluated using a satellite semimajor axis from 0.25 to 0.5 R$_{\rm H}$ in steps of 0.001 R$_{\rm H}$, where R$_{\rm H}$ is the planet's Hill radius {and the range in R$_{\rm H}$ is motivated by previous observational and dynamical studies of satellites \citep{Cruikshank1982,Saha1993,Jewitt2007,Domingos2006,Donnison2010,Rosario-Franco2020,Quarles2021}}.  Many studies \citep{Henon1970,Hamilton1991,Morais2012,Grishin2017,Quarles2021} have demonstrated that retrograde orbital stability extends to larger values of the satellite semimajor axis compared to the prograde case.  Thus, we increased the $a_{\rm sat}$ range to 0.45--0.70 R$_{\rm H}$ with a 0.001 R$_{\rm H}$ step size.  In physical units, a 0.001 R$_{\rm H}$ step corresponds to approximately 0.0001 au, noting that we scale the steps with respect to the Hill radius as this approach will allow our results to scale with improved characterizations of the planetary and stellar parameters, if available.

{In each simulation, the} moon begins on a circular orbit that is apsidally aligned ($\omega_{\rm sat}=\Omega_{\rm sat}=0^\circ$) with the planetary orbit. For each combination of planetary eccentricity and the satellite's semimajor axis, 20 simulations are evolved using a random mean anomaly $\theta$ for the satellite chosen from 0$^\circ$--$359^\circ$.  We use a parameter $f_{\rm stab}$ to summarize these trials, which represents the fraction of stable simulations for a given ($e_{\rm p}, a_{\rm sat}$) combination.  

\subsection{Satellite orbital migration due to tides}

A satellite's long-term evolution is affected by tides raised on its host planet, where the induced tidal bulge slows the planet's rotation over billion-year timescales.  Through the conservation of angular momentum, the satellite can fall toward the planet or migrate outward toward the Hill radius.  The satellite's migration depends on whether its orbital period $T_{\rm sat}$ is greater than (outward migration) or less than (inward migration) the host planet's rotation period $P_{\rm rot}$.  We begin the satellite on a circular, coplanar orbit relative to the Roche radius ($a_{\rm sat}=3R_{\rm roche}$) {with the satellite treated as a fluid satellite, and Roche radius calculated via}:

\begin{equation}
    R_{\rm roche} \approx 2.44 R_{\rm p}(\rho_{\rm p}/\rho_{\rm sat})^{1/3},
\end{equation}
where the planet radius $R_{\rm p}$ {is assumed to equal the radius of Jupiter given the well-established trends in the mass radius relation for giant planets \citep{Fortney2007,Chen2017}}, the planet density $\rho_{\rm p}$ is 1.33 g~cm$^{-3}$ (Jupiter-like), and the satellite density is 5.515 g~cm$^{-3}$ (Earth-like).  {A satellite at 3$\times$ the Roche radius begins an orbital period of $\sim$18 hr}.  From theoretical calculations and numerical simulations of giant planet formation \citep{Takata1996,Batygin2018}, giant planets are expected to be rapid rotators ($\sim$3 hr) due to gas accretion or rotate more slowly ($\sim$10--12 hr) {like the Solar System giant planets} if magnetic braking is efficient.  Since the satellite's orbital period ($\sim$18 hr at $3 R_{\rm roche}$) is greater than the expected spin period of giant planets, the satellite will undergo outward migration.

Equilibrium tidal models are commonly prescribed within two types: constant phase lag (CPL; \citealt{Goldreich1966}) or constant time lag (CTL; \citealt{Hut1981}).  Both models require an assumption for the Love number $k_2$ \citep{Love1911} and a moment of inertia factor $\alpha$, for which we use Jupiter-like values (0.565 and 0.2756, respectively) determined from the \textit{Juno} probe \citep{Ni2018,Idini2021}.  The tidal models differ in their approach to approximating the tidal dissipation $\epsilon$, where the CPL model implements a constant $Q$ and the CTL model uses a constant timelag $\tau$.

Both models yield similar results for small satellite-planet mass ratios, but the CTL model more accurately represents the tidal forcing frequencies \citep{Ogilvie2014}; thus, we use a CTL model.  The constant timelag $\tau$ is unknown for HD 23079b; hence, we assume a Jupiter-like value ($\tau_{\rm J}\sim 0.035$ s) while evaluating models over several orders of magnitude ($10^{-2}-10^2\;\tau_{\rm J}$).  The planetary mass given in Table \ref{tab:sys_param} is determined through the RV method, which allows an observer to determine the minimum mass $m_{\rm p}$.  Therefore, we evolve the tidal model considering three host planet masses (1, 1.5, and 2 $m_{\rm p}$) with a Jupiter-like $\tau$.  Herein we use the CTL model derived by \citet{Hut1981} assuming zero planetary obliquity, which is equivalent to the formalism described in more recent approaches  \citep{Leconte2010,Heller2011,Barnes2017}.

The tidal evolution {with respect to time $t$} is described by the following equations:
\begin{align}
\frac{da_i}{dt} &= \frac{2a_i^2 Z_{{\rm p},j}}{G m_{\rm p} m_j}  \left(\frac{f_2(e_i)}{\beta^{12}(e_i)} \frac{\Omega_{\rm p}}{n_j} - \frac{f_1(e_i)}{\beta^{15}(e_i)} \right), \label{eqn:a_i}\\
\frac{de_i}{dt} &= \frac{11a_ie_i Z_{{\rm p},j}}{2G m_{\rm p} m_j} \left(\frac{f_4(e_i)}{\beta^{12}(e_i)} \frac{\Omega_{\rm p}}{n_j} - \frac{18}{11}\frac{f_3(e_i)}{\beta^{13}(e_i)} \right), \label{eqn:e_i}
\end{align}
\noindent and
\begin{align} \label{eqn:spin}
\frac{d\Omega_{\rm p}}{dt} = \sum_j \frac{Z_{{\rm p},j}}{2 \alpha_{\rm p} m_{\rm p}  R^2_{\rm p} n_j} \left
(\frac{2 f_2(e_j)}{\beta^{12}(e_j)} - \frac{f_5(e_j)}{\beta^9(e_j)} \frac{\Omega_{\rm p}}{n_j}
\right),
\end{align}
\noindent where
\begin{equation}
Z_{{\rm p},j} \equiv 3G^2k_{2,\rm p}\tau_{\rm p} m^2_j ( m_{\rm p} + m_j ) \frac{R_{\rm p}^5}{a_i^9} 
\end{equation}
\noindent and
\begin{equation}
    \begin{aligned}
    \beta(e) &= \sqrt{1 - e^2}, \\
    f_1(e) &= 1 + \frac{31}{2}e^2 + \frac{255}{8}e^4 + \frac{185}{16}e^6 + \frac{25}{64}e^8, \\
    f_2(e) &= 1 + \frac{15}{2}e^2 + \frac{45}{8}e^4 + \frac{5}{16}e^6 \\
    f_3(e) &= 1 + \frac{15}{4}e^2 + \frac{15}{8}e^4 + \frac{5}{64}e^6, \\
    f_4(e) &= 1 + \frac{3}{2}e^2 + \frac{1}{8}e^4,\\
    f_5(e) &= 1 + 3e^2 + \frac{3}{8}e^4.
\end{aligned}
\end{equation}
\noindent Equations \ref{eqn:a_i} and \ref{eqn:e_i} describe the semimajor axis and eccentricity evolution of either the planet or satellite through the subscript $i$.  The subscript $j$ represents either the host star or the satellite that is raising the tide on the planet, where $n_j$ is the respective orbital mean motion.  Equation \ref{eqn:spin} describes the spin evolution of the planet, where the moon is assumed to be synchronously rotating and the changes to the host star's spin is negligible.  The subscript $j$ in Eqn. \ref{eqn:spin} represents either the host star or the satellite that is contributing to spin-down the planet.  A Jupiter-like value is used for the moment of inertia factor ($\alpha_{\rm p}=0.565$) {and $G$ represents the Newtonian constant of gravitation}.


%
\begin{table*}
\begin{center}
	\caption{Stellar and planetary parameters \label{tab:sys_param}}
\begin{tabular}{lcl}
\noalign{\smallskip}
\hline
\noalign{\smallskip}
Parameter$^{a}$ & Value & Reference  \\
\noalign{\smallskip}
\hline
\noalign{\smallskip}
Spectral type                    & F9.5V              &
\cite{Gray2006} \\
RA                               &
03$^{\rm h}$ 39$^{\rm m}$ 43.0961$^{\rm s}$           &
\cite{Gaia2018} \\
DEC                              &
${-}52^\circ$ 54$^\prime$ 57.0161$^{\prime\prime}$    & \cite{Gaia2018} \\
Apparent magnitude V             & 7.12               &
\cite{Anderson2012} \\
Distance (pc)                    & 33.49 $\pm$ 0.03   &
\cite{Anderson2012} \\
$M$ (M$_\odot$)                  & 1.01 $\pm$ 0.02    &
\cite{Bonfanti2015} \\
$T_{\rm eff}$ (K)                & 6003 $\pm$ 36      &
\cite{Bonfanti2015} \\
$R$ (R$_\odot$)                  & 1.08 $\pm$ 0.02    &
\cite{Bonfanti2015} \\
$L$ (L$_\odot$)                  & 1.372 $\pm$ 0.005  &
\cite{Bonfanti2015} \\
Age (Gyr)                        & 5.1 $\pm$ 1.0      &
\cite{Bonfanti2015} \\
$m_{\rm p}{\sin}\;i$ (M$_{\rm J}$) & 2.41 $\pm$ 0.6   &  \cite{Wittenmyer2020} \\
$P$ (days)                       & 724.5 $\pm$ 2.2    &  \cite{Wittenmyer2020} \\
$a_{\rm p}$ (au)                 & 1.586 $\pm$ 0.003  & \cite{Wittenmyer2020}  \\
$e_{\rm p}$                      & 0.087 $\pm$ 0.031  &
\cite{Wittenmyer2020}  \\
\noalign{\smallskip}
\hline
\end{tabular}
\vspace{0.05in}
\begin{enumerate}
\item[$^a$] All parameters and symbols have their customary meaning.
\end{enumerate}
\end{center}
\end{table*}

%
\begin{table*}
	\centering
	\caption{$N$-body simulation parameters}
\begin{tabular}{lcl}
\noalign{\smallskip}
\hline
\noalign{\smallskip}
Parameter & Range & Step \\
\noalign{\smallskip}
\hline
\noalign{\smallskip}
$a_{\rm p}$ & 1.586 au & fixed \\
$e_{\rm p}$ & 0.05--0.12 & 0.001 \\
$a_{\rm sat}$ ($i_{\rm sat} = 0^\circ$) & 0.25--0.50 R$_{\rm H}$ & 0.001 R$_{\rm H}$\\
$a_{\rm sat}$ ($i_{\rm sat} = 180^\circ$) & 0.45--0.70 R$_{\rm H}$ & 0.001 R$_{\rm H}$\\
$\theta_{\rm sat}$ & 0$^\circ$--359$^\circ$ & random \\
\noalign{\smallskip}
\hline
\end{tabular}
\end{table*}

%

\section{RESULTS and DISCUSSION}

\subsection{Model simulations}
Recent observations by \citet{Benisty2021} {revealed the existence of a circumplanetary disk around PDS 70c, a planet observed to be in the process of accreting gas}.  After this stage, more massive satellites could be acquired through processes of tidal capture and pull down \citep{Hamers2018} as has been suggested for the candidate exomoon Kepler 1625b-I \citep{Teachey2018}.  {Assuming that exomoons form soon after the epoch of planet formation,} such moons must survive against perturbations from the host star to be observed in the present day ($\sim$5 Gyr; \cite{Bonfanti2015}). Our goal is to determine the stability boundary of putative satellites around the host planet, HD 23079b. Previously, \citet{Eberle2011} and \citet{Cuntz2013} discussed the orbital stability limit of an Earth-mass object in this system as a Trojan planet or a natural satellite within the planet's Hill radius.

We examine the orbital stability limit for prograde and retrograde orbits using $N$-body simulations with \texttt{REBOUND} (see Sect. \ref{sec:nbody}).  These simulations consider a range of initial planetary eccentricity consistent with current observational constraints \citep{Wittenmyer2020}.  \citet{Rosario-Franco2020} and \citet{Quarles2021} provided rough estimates for the stability limit in terms of the planet's Hill radius, whereas here we explore this system in much finer detail.  Similar to \citet{Rosario-Franco2020} and \citet{Quarles2021}, we use the lower critical orbit \citep{Rabl1988} to define the stability limit, which is a more conservative approach that excludes regions of quasi-stability.

%
\begin{figure*}
\centering
  \includegraphics[width=0.8\linewidth]{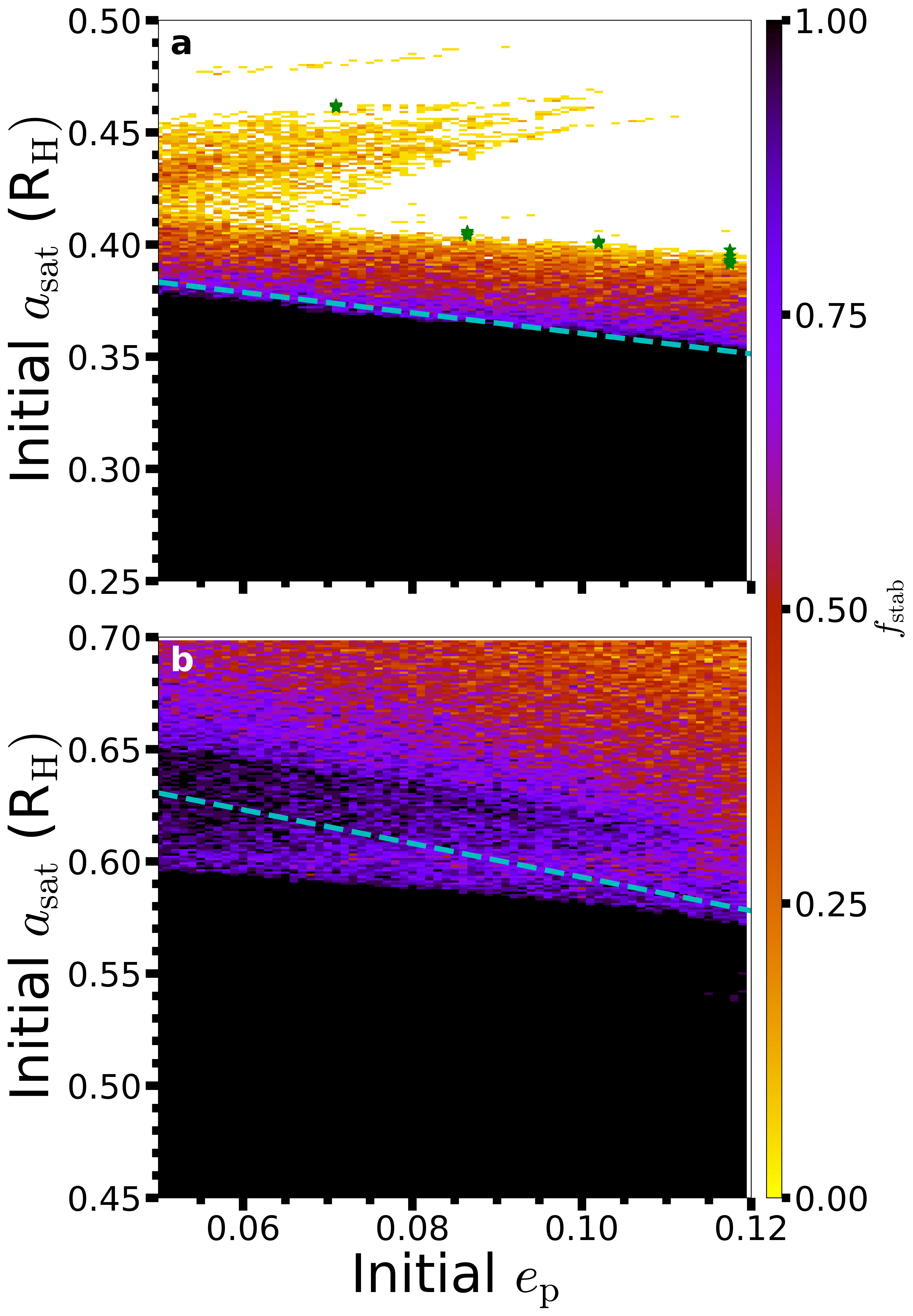}
  \caption{Numerical estimates for the stability of (a) prograde and (b) retrograde exomoons orbiting HD 23079b as a function of the satellite's initial semimajor axis $a_{\rm sat}$ in units of the planetary Hill radius R$_{\rm H}$ and the planetary eccentricity $e_{\rm p}$.  The color code represents the fraction $f_{\rm stab}$ (out of 20) of stable simulations for a $10^5$ yr timescale; it shows which initial parameters depend on the initial placement of the satellite through its mean anomaly $\theta_{\rm sat}$.  The white cells denote cases where zero trial simulations survive for $10^5$ yr and, conversely, the black cells denote cases where all the trial simulations survive.  The cyan (dashed) lines mark the expected stability limits for (a) prograde \citep{Rosario-Franco2020} and (b) retrograde \citep{Quarles2021} orbiting exommons.  The green stars in (a) mark the previous estimates from \cite{Cuntz2013}, which are found to lie at the border of the quasi-stable regime. \label{fig:stab_map} }
\end{figure*}

Figure \ref{fig:stab_map} demonstrates the results of our simulations in terms of initial semimajor axis of the satellite $a_{\rm sat}$ (in R$_{\rm H}$) and the planetary eccentricity $e_{\rm p}$. Figures \ref{fig:stab_map}a and \ref{fig:stab_map}b are color-coded using the parameter $f_{\rm stab}$, which is defined as the fraction of 20 simulations with random mean anomalies for the satellite that survive for $10^5$ yr. The values of $f_{\rm stab}$ range from 0.0--1.0, where the cells with $f_{\rm stab}<0.05$ (wholly unstable) are colored white.  The fully stable ($f_{\rm stab}=1$; black) cells are used in our calculation of the stability boundary, while the {values} between the extremes illustrate regions of quasi-stability.  The dashed (cyan) curves mark the stability limit previously determined for prograde \citep{Rosario-Franco2020} and retrograde \citep{Quarles2021} satellites, respectively.

For prograde orbits (Fig. \ref{fig:stab_map}a), the stability limit extends to 0.37 R$_{\rm H}$ for the lowest consider planet eccentricity and decreases to 0.35 R$_{\rm H}$ for larger planetary eccentricity. Our stability limit closely agrees with the stability fitting formula by \citet{Rosario-Franco2020}.  Beyond this boundary, there is a gradient of quasi-stability over a small range in satellite semimajor axis.  At $\sim$0.43 R$_{\rm H}$, there is a 6:1 (first-order) mean motion resonance (MMR) between the planet and satellite orbits \citep{Quarles2021}.  The MMR excites the satellite's eccentricity over time, which allows for the satellite to escape as its apocenter extends beyond the upper critical orbit ($\approx$ 0.5 R$_{\rm H}$; \citet{Domingos2006}).  The MMR's resonant angle depends on the relative orientation (i.e., mean anomaly) of the planetary and satellite orbits and particular starting angles can survive for longer periods, if the satellite returns to approximately the same phase after six orbits.  \citet{Cuntz2013} used a single initial mean anomaly for the satellite, which largely corresponds to the upper critical orbit (green stars in Fig. \ref{fig:stab_map}a). 

For retrograde orbits (Fig. \ref{fig:stab_map}b), the continuously stable region extends to $\sim$0.59 R$_{\rm H}$ for $e_{\rm p}$= 0.05 and recedes to $\sim$0.57 R$_{\rm H}$ for $e_{\rm p}$ = 0.12 in a similar manner as the stability limit for Fig. \ref{fig:stab_map}a. For the initial planetary eccentricity from 0.05 to 0.10, there is a stable peninsula corresponding to a 7:2 (second-order) MMR. As MMRs increase in order, the magnitude of the eccentricity excitation decreases \citep{Murray1999}. Moreover, the weakened Coriolis force and shorter interaction times for retrograde orbits \citep{Henon1970} also reduce the magnitude of secular eccentricity excitation.  \citet{Quarles2021} considered a coarser grid of simulations, which did not resolve the gap created by the 4:1 (first-order) MMR.  Hence, the stability limit was slightly over-estimated (dashed curve) in their work.  However, it is a better approximation of the stability limit compared to previous works that focused on the upper critical orbit \citep{Domingos2006,Cuntz2013}.  The limits for retrograde orbits from \citet{Cuntz2013} are larger than 0.7 R$_{\rm H}$ as are those by \citet{Domingos2006}.  Hence, the previous results from \cite{Cuntz2013} are not shown in Fig. \ref{fig:stab_map}b.

Available mass measurements of HD 23079b are based on the RV method; thus, only the minimum mass is known implying the true mass of HD 23079b could be higher.  {Generally, we expect the true mass to differ by a factor of 1/$\sin(\pi/4)$ (or $\sim$1.4) assuming an isotropic distribution restricted to prograde orbits for the planetary inclination on the sky plane.}  In Fig. \ref{fig:stab_map}, we use the minimum mass $m_{\rm p}$ in all our calculations.  Estimates of the exoplanet mass distribution \citep{Jorissen2001,Ananyeva2020} indicate that planets with a substantially increased mass are rare and thus we expect the true mass of HD 23079b to differ from the minimum mass by only a small factor.  Hence, we perform another set of stability simulations for prograde moons with the host planet's mass is increased to 1.5 $m_{\rm p}$ (i.e., 3.62 M$_{\rm J}$).

\begin{figure*}
\centering
  \includegraphics[width=\linewidth]{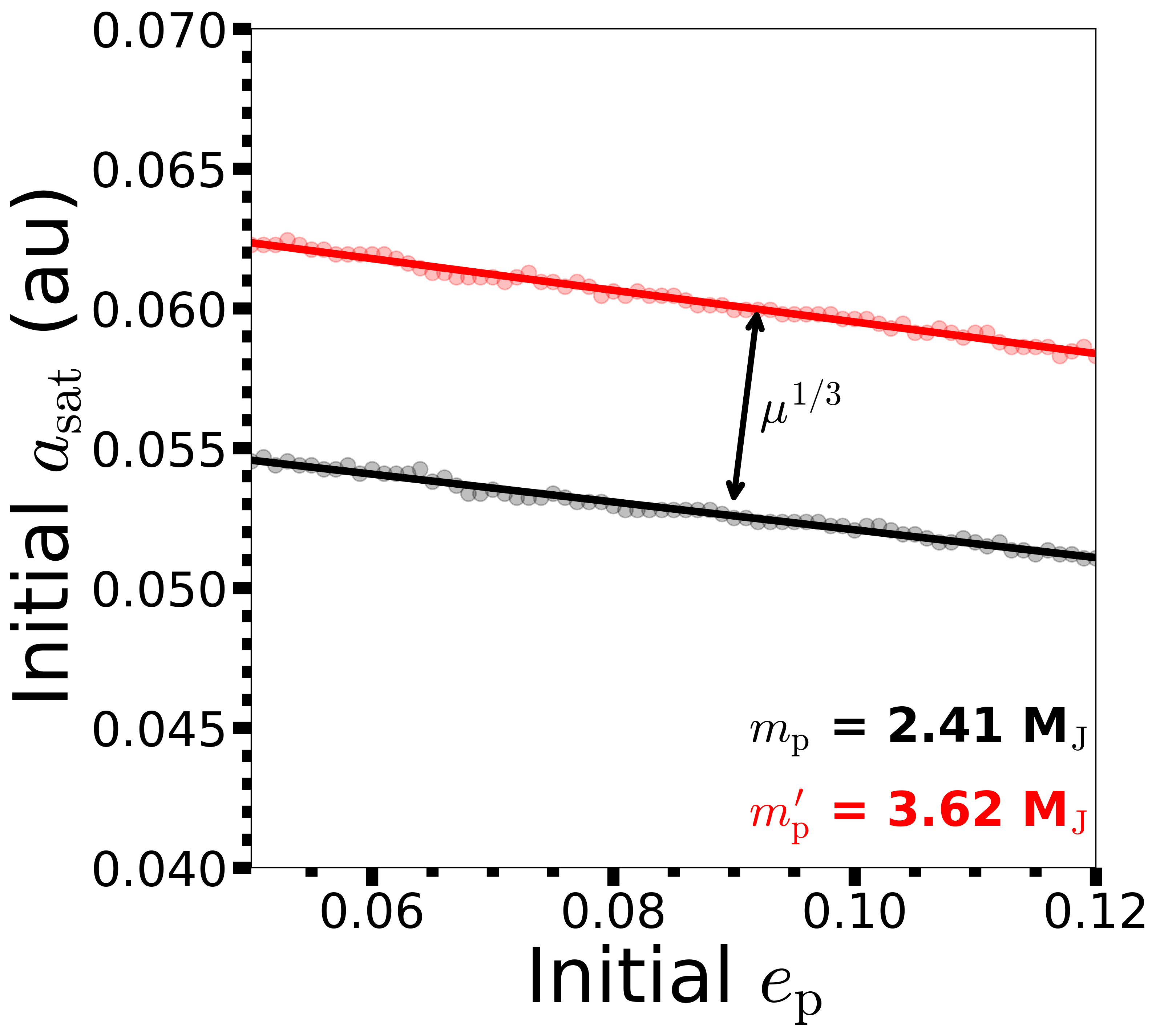}
  \caption{Stability limits for prograde exomoons assuming the minimum planet mass of $m_{\rm p} = 2.41~{\rm M}_{\rm J}$ (black) and an increased mass of $m_{\rm p}^{\prime} = 3.62~{\rm M}_{\rm J}$ (red).  The best-fit curves scale as a power law with the mass ratio $\mu = m_{\rm p}^{\prime}/m_{\rm p}$. Note that the $y$-axis values are in physical units (au) instead of R$_{\rm H}$. \label{fig:stab_scale} }
\end{figure*}

Figure \ref{fig:stab_scale} shows the stability limit (in au) for the minimum mass $m_{\rm p}$ (black) and the increased mass $m_{\rm p}^\prime$ (red) as a function of the planetary eccentricity $e_{\rm p}$.  The stability limit (in au) clearly increases for a larger planet mass because the respective Hill radius is larger (see Eqn. \ref{eqn:R_H}).  Thus, we expect the red curve to scale by a factor $\mu = m_{\rm p}^\prime/m_{\rm p}$.  Comparing the two curves (black and red) indicates that the stability limit increases by a factor of $\sim$1.14 (i.e., ${1.5^{1/3}}$).  If future observations reveal a planetary mass beyond the minimum value, the stability limits as obtained can be readily adjusted through a simple scale factor \citep{Wittenmyer2020}.



%

 \subsection{Tidal migration}
 The orbits of natural satellites (including our Moon) have migrated since the time of their formation due to de-spinning of their host planet from tides raised from the Sun and the satellites {\citep{Goldreich1966,Goldreich1966b,Touma1998,Cuk2012}}.  We evaluate the possible extent of migration for a putative Earth-mass moon orbiting HD 23079b using Eqns. \ref{eqn:a_i}--\ref{eqn:spin}, which describe the tidal migration based on the CTL model \citep{Hut1981,Barnes2017}. The satellite begins on a circular orbit at $3 R_{\rm roche}$ (or $\approx$ 0.015 R$_{\rm H}$), where the initial planetary rotation period is varied from 3 to 12 hr in 0.25 hr steps.  \citet{Piro2018} showed that the satellite's semimajor axis after 10 Gyr can differ depending on the assumed planetary rotation rate. To test this dependence on the assumed $e_{\rm p}$, we consider a range of values from 0.05 to 0.13 in steps of 0.01.
 
%
\begin{figure*}
\centering
  \includegraphics[width=\linewidth]{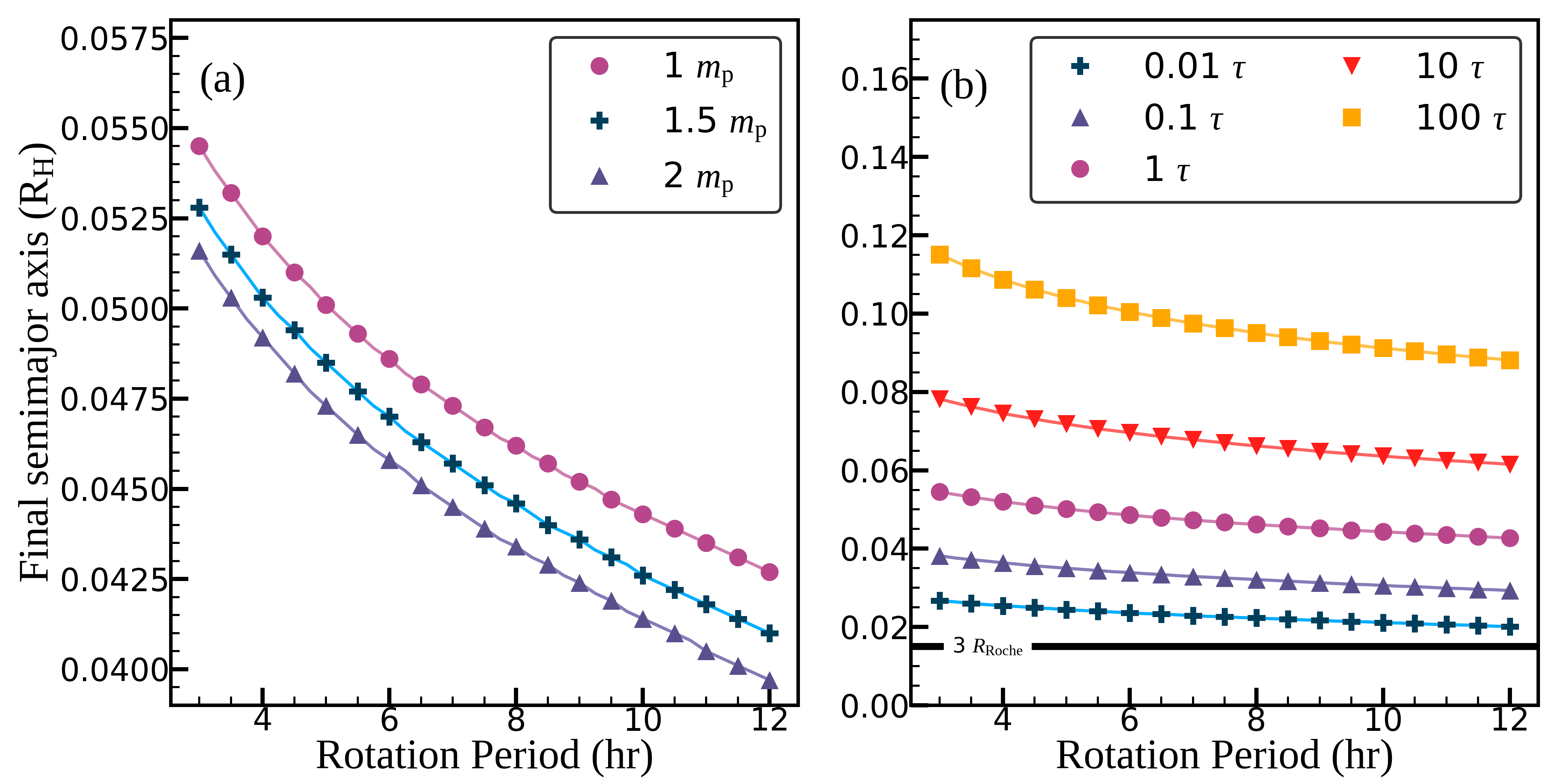}
  \caption{Relationship between the satellite's final semimajor axis and the planetary rotation period based on tidal model simulations for different values of $m_{\rm p}$ and $\tau$. (a) Results varying the assumed planetary mass (1.0 $m_{\rm p}$, 1.5 $m_{\rm p}$, and 2.0 $m_{\rm p}$) while using a Jupiter-like tidal time lag $\tau$. (b) Results varying the tidal time lag from 0.01 $\tau$ to 100 $\tau$. The horizontal line in (b) at 0.015 R$_{\rm H}$ represents $3 R_{\mathrm{Roche}}$. Note the difference in the $y$-axis ranges between panel (a) and (b). Additionally, the 1 $m_{\rm p}$ curve in (a) and the 1 $\tau$ cure in (b) are identical. In panel (b), only every other calculation has been depicted by a marker for increased clarity of the figure. \label{fig:tides}}
\end{figure*}

From our calculations, we find that there were no notable changes in the final semimajor axis for all $e_{\rm p}$ values considered in this study. This is because the host planet is not close to the star and thus the stellar tides are largely negligible.  However, the host star also has a larger influence on the satellite's orbit and impacts the satellite's eccentricity.  This kind of forcing depends on the semimajor axis ratio ($a_{\rm sat}$/$a_{\rm p}$) and the planetary eccentricity  ($e_{\rm p}$/(1-$e_{\rm p}^2)$) \citep[e.g.,][]{Heppenheimer1978,Andrade-Ines2017}; both of which are very small.  Since the moon's forced eccentricity is small, the eccentricity contribution to the star-planet and planet-moon tides is also small. Therefore, we present results that only use $e_{\rm p}$ = 0.09 in our simulations.

Figure \ref{fig:tides} demonstrates the final semimajor axis of the satellite as a function of the assumed planetary rotation period due to the tidal evolution over 10 Gyr. We use the observationally determined minimum mass of the planet ($m_{\rm p}$ = 2.41 M$_{\rm J}$) and a Jupiter-like constant time lag ($\tau_{\rm p}$ = $\tau_{\rm J}$) for HD 23079b. The magenta line (with dots) represents this nominal case in Figs. \ref{fig:tides}a and \ref{fig:tides}b. The final semimajor axis of the satellite under our nominal conditions is $\sim$0.0545 R$_{\rm H}$ for the fastest rotation period (3 hr) and $\sim$0.043 R$_{\rm H}$ for the slowest rotation period (12 hr).  Observations from the RV method restrict the planetary mass measurement to the minimum mass, a limitation that could be overcome in the future.  We analyze several other cases that vary the assumed planetary mass by a factor of 1.5 and 2 (see Fig. \ref{fig:tides}a). The satellite's final semimajor axis $a_{\rm fin}$ decreases for a larger planetary mass (relative to minimum mass $m_{\rm p}$) by the mass ratio $\mu$ $=m_{\rm p}^\prime/m_{\rm p}$, which scales by a power law; i.e., $a_{\rm fin}^\prime \propto \mu^{-1/12}a_{\rm fin}(P_{\rm rot})$.  

In Fig. \ref{fig:tides}b, we vary the dissipation strength through the constant time lag over four orders of magnitude ($C_\tau = 10^{-2}-10^{2}~\tau$).  Interestingly, a 100 fold increase in $\tau$ (orange squares in Fig. \ref{fig:tides}b) results in a doubling of the final satellite semimajor axis when the planetary rotation period $P_{\rm rot}$ is 3 hr; i.e., $a_{\rm fin}^\prime \propto C_\tau^{1/6}a_{\rm fin}(P_{\rm rot})$.  The satellite's semimajor axis evolution (see Eqn. \ref{eqn:a_i}) depends linearly on the assumed value for $\tau_{\rm p}$, but it also depends non-linearly on $\tau_{\rm p}$ through the changes in the planetary rotation rate $\Omega_{\rm p}$ (see Eqn. \ref{eqn:spin}).  The combination of those dependencies are the likely underlying cause of the empirically derived scaling relation.

Irregardless of the assumed parameters for the tidal evolution, the final $a_{\rm sat}$ is far from the stability limit, where the largest final $a_{\rm sat}$ is only $\sim$1/3 of the prograde stability limit. The tidal force is known to decrease rapidly with distance.  Thus, starting the satellite at most separations would not affect the satellite's potential stability \citep{Quarles2020b}.  However, the CTL model considers tidal migration secularly without any interruptions due to MMRs or changes in the internal evolution of the host planet {\citep{Touma1998}}. Realistically, such interactions could include slowing down the migration process temporarily.  In fact, this kind of behavior may have occurred for our Moon \citep{Sasaki2012}.  {Tidal evolution with multiple moons could also induce some volcanic activity as is the case for Io \citep{Peale1979} and potentially affect an exomoon's habitability \citep{Heller2013}, but such considerations are beyond the scope of this work.}

\subsection{Observability of possible exomoons in the HD~23079 system}
 {The detection of exomoons is currently extremely challenging but their detection is technically feasible}, where \citet{Sartoretti1999} showed the transit method as a promising avenue for their eventual discovery.  A dedicated search for exomoons within the {\it Kepler} data \citep{Kipping2012,Kipping2013a, Kipping2013b,Kipping2014,Kipping2015b} has yet to confirm an exomoon, while noting that Kepler 1625b-I represents an interesting candidate \citep{Teachey2018}. To observe an exomoon in the HD 23079 system, a different approach is required since the host planet was discovered through the RV method \citep{Tinney2002,Wittenmyer2020} and is not known to transit its host star relative to our line-of-sight.  The expected semi-amplitude $K_{\rm o}$ from the stellar motion about the center-of-mass is $\sim$54 m~s$^{-1}$, where the addition of an Earth-mass satellite orbiting HD 23079b would introduce a small additional variation ($<$1 m~s$^{-1}$).  Consequently, the most promising technique is Doppler monitoring within direct imaging observations \citep{Vanderburg2018}, where an RV signal is extracted from the host planet's reflex motion after accounting for variations in the host planet's reflected light.

\begin{figure*}
\centering
  \includegraphics[width=\linewidth]{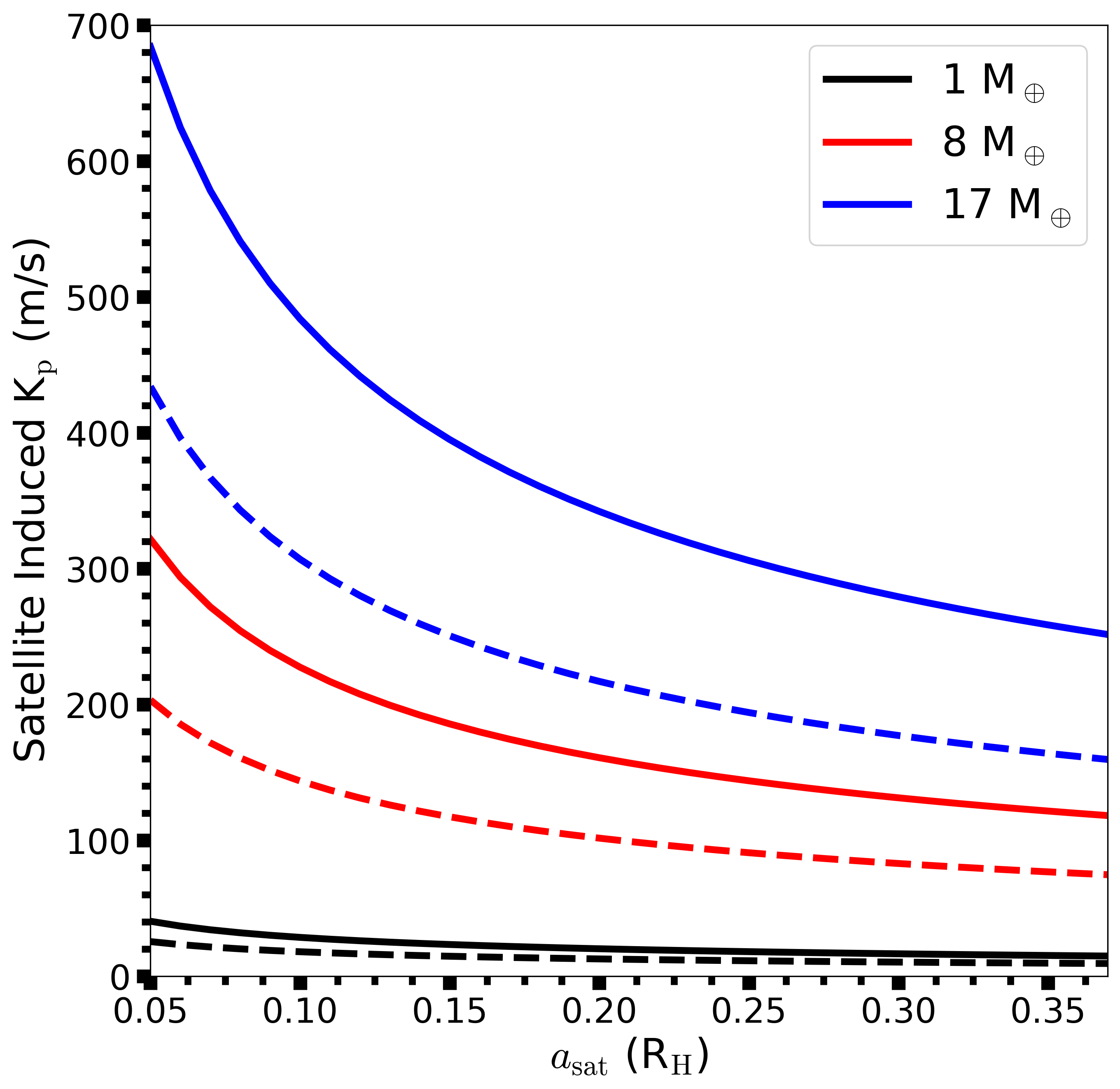}
  \caption{The RV semi-amplitude $K_{\rm p}$ induced by a satellite with respect to the the planet-satellite semimajor axis $a_{\rm sat}$ in units of the host planet's Hill radius R$_{\rm H}$.  The solid curves represent values assuming the minimum mass ($m_{\rm p} = 2.41 {\rm M}_{\rm J}$) is the true planetary mass.  The dashed curves illustrate the reduction in $K_{\rm p}$ for double the minimum mass (2 $m_{\rm p}$).  The curves are color-coded (black, red, and blue) to mark the differences in the assumed satellite-mass (1, 8, and 17 M$_\oplus$, respectively). \label{fig:RVs}}
\end{figure*}

The host planet's semi-amplitude $K_{\rm p}$ induced by an exomoon  \citep{Vanderburg2018,Perryman2018} is given by the following:
\begin{equation}
    K_{\rm p} = \left(\frac{m_{\rm sat}\sin{i_{\rm sat}}}{m_{\rm p} + m_{\rm sat}}\right) \sqrt{\frac{Gm_{\rm sat}}{a_{\rm sat}\left(1-e_{\rm sat}^2\right)}},
\end{equation}
where the satellite orbital inclination $i_{\rm sat}$ is relative to the observer's line-of-sight and should be similar in magnitude to the observed planetary inclination due to tidal evolution of the planet-satellite pair \citep{Porter2011}.  Although the system is not known to transit, we assume that $i_{\rm sat}=90^\circ$ to estimate the maximum $K_{\rm p}$.  Figure \ref{fig:RVs} demonstrates the maximum satellite induced $K_{\rm p}$ as a function of the satellite semimajor axis $a_{\rm sat}$ in units of R$_{\rm H}$, where the reflex velocity on the planet's orbit about the barycenter decreases as the satellite semimajor axis increases ($K_{\rm p} \propto a_{\rm sat}^{-1/2}$).  The black, red, and blue solid curves mark when an Earth-mass, a standard super-Earth (8 M$_\oplus$), or a Neptune (17 M$_\oplus$), respectively, is assumed for the satellite and the RV minimum mass ($m_{\rm p}=2.41~{\rm M}_{\rm J}$) is used.  The dashed curves are provided to show how much the satellite-induced RV signal decreases, if the assumed planetary mass is doubled (2$m_{\rm p}$).

Figure \ref{fig:RVs} shows that massive ($\gtrsim$ 8 M$_\oplus$), prograde-orbiting satellites could produce a Keplerian signal with an RV semi-amplitude greater than $\sim$100 m~s$^{-1}$, even with a satellite semimajor axis near the stability limit.  Keplerian signals from prograde, Earth-mass satellites are limited to $\sim$20--40 m~s$^{-1}$.  Retrograde-orbiting satellites stably orbit at larger separations with $a_{\rm sat} \leq 0.59\:{\rm R}_{\rm H}$, but the resulting Keplerian signal would be less optimal for the observability.

The current best RV precision is $\sim$1 m~s$^{-1}$, where this precision level is only attainable for bright (V$<10$) stars.  The host star in HD 23079 is relatively bright (V=7.12; \citet{Anderson2012}), but the direct imaging method proposed by \citet{Vanderburg2018} would allow {the analysis of} the much fainter reflected light from the planet, which would be much more limited in precision ($\sim$1,500 m~s$^{-1}$).  However, large {(30 m class)} telescopes (e.g., Giant Magellan Telescope; \citet{Jaffe2016}) are on the horizon and should be available in the foreseeable future.
They would make Doppler surveys of directly imaged planets attainable due to the much larger S/N compared to current technology affecting the RV precision \citep{Quanz2015}.  In particular, the {detection of} Keplerian signals from massive exomoons with an RV semi-amplitude greater than $\sim$100 m~s$^{-1}$ would be feasible.

%

\section{SUMMARY AND CONCLUSIONS}
The aim of our study is to further explore the possibility of exomoons in the HD~23079 system.  In this system, a solar-type star of spectral {class} F9.5V hosts a Jupiter-mass planet in a nearly circular orbit
situated in the outer segment of the stellar habitable zone.  Previous studies have examined the orbital stability limit of an Earth-mass object in this system as a Trojan planet \citep{Eberle2011} or a natural satellite \citep{Cuntz2013}.  We focus on the {latter} to more accurately identify the stability limits for prograde and retrograde exomoons within observational constraints, including the recent work by \citet{Wittenmyer2020}.  

In the past year, \citet{Rosario-Franco2020} updated the fitting formulas for the stability limit for prograde-orbiting satellites in terms of the planet's Hill radius, whereas \cite{Quarles2021} improved the fitting formulas for retrograde systems.  We follow the prior approaches, in much finer detail, for the HD~23079 system, where the stability limits determined herein specifically exclude regions of quasi-stability and resonances.  Additionally, we evaluate multiple satellite mean anomalies, which allows us to overcome some limitations from previous works
\citep[e.g.,][]{Domingos2006}.

Our study shows that the system of HD~23079 is a highly promising candidate for hosting {potentially} habitable exomoons despite the fact that the outer stability limit is modestly reduced.  Noting that HD~23079b's mass is not exactly known --- as due to the RV detection technique only a minimum value could hitherto been identified --- our results are still applicable, if a more precise mass value becomes available as the outer orbital stability limit follows a well-defined scaling law, i.e., $(m_{\rm p}^\prime/m_{\rm p})^{1/3}$; see text for details.
The outward migration due to tides does not greatly affect the potential stability of exomoons in a CTL tidal model \citep{Hut1981,Barnes2017}, where we find that a putative satellite's migration distance the stellar lifetime scales inversely to the 1/12th power in mass ratio $\mu$ when comparing different assumptions on planetary mass from the sky plane inclination.  Moreover, we find that migration distance scales inversely to the 1/6th power in the assumed tidal time lag parameter $\tau$ relative to a Jupiter-like value.  Scaling relations, in either the mass or tidal time lag, would assist in the general search for exomoons as well as future observations of the HD~23079 system.

We also explore the observability of putative HD~23079 exomoons.  Current technologies are incapable of identifying moons in that system; however, future developments hold promise.  As the transit method is unavailable for finding exomoons in HD~23079, Doppler monitoring within direct imaging observations might offer positive outcomes.  Note that large {(30 m class)} telescopes (e.g., Giant Magellan Telescope; \citet{Jaffe2016}) should be available in the foreseeable future.  The much larger S/N from telescopes with a large mirror would make Doppler surveys of directly imaged planets attainable{, where} the Keplerian signals from Earth-mass exomoons with an RV semi-amplitude greater than $\sim$100 m~s$^{-1}$ would be possible \citep{Vanderburg2018}.

\begin{acknowledgements}
This research was supported in part through research cyberinfrastructure resources and services provided by the Partnership for an Advanced Computing Environment (PACE) at the Georgia Institute of Technology.  {The authors thank the anonymous reviewer for comments that helped improve the quality and clarity of the manuscript.} 
\end{acknowledgements}

\bibliographystyle{pasa-mnras}
\bibliography{references}

\end{document}